# Alternation of Magnetic Anisotropy Accompanied by Metal-Insulator Transition in Strained Ultrathin Manganite Heterostructures

Masaki Kobayashi[1,2,*], Le Duc Anh[1,3,4], Masahiro Suzuki[5], Shingo Kaneta-Takada[1], Yukiharu Takeda[6], Shin-ichi Fujimori[6], Masaaki Tanaka[1,2], Shinobu Ohya[2,3], and Atsushi Fujimori[5,6,7]

[1]*Department of Electrical Engineering and Information Systems, The University of Tokyo, 7-3-1 Hongo, Bunkyo-ku, Tokyo 113-8656, Japan*

[2]*Center for Spintronics Research Network, The University of Tokyo, 7-3-1 Hongo, Bunkyo-ku, Tokyo 113-8656, Japan*

[3]*Institute of Engineering Innovation, Graduate School of Engineering, The University of Tokyo, 7-3-1 Hongo, Bunkyo-ku, Tokyo 113-8656, Japan*

[4]*PRESTO, JST, 4-1-8 Honcho,k Kawaguchi, Saitama, 332-0012, Japan*

[5]*Department of Physics, The University of Tokyo, 7-3-1 Hongo, Bunkyo-ku, Tokyo 113-0033, Japan*

[6]*Materials Sciences Research Center, Japan Atomic Energy Agency, Sayo-gun, Hyogo 679-5148, Japan*

[7]*Department of Applied Physics, Waseda University, Okubo, Shinjuku, Tokyo 169-8555, Japan*

*Email: masaki.kobayashi@ee.t.u-tokyo.ac.jp

**Abstract**

Fundamental understanding of interfacial magnetic properties in ferromagnetic heterostructures is essential to utilize ferromagnetic materials for spintronic device applications. In this paper, we investigate the interfacial magnetic and electronic structures of epitaxial single-crystalline $LaAlO_3$ (LAO)/$La_{0.6}Sr_{0.4}MnO_3$ (LSMO)/Nb:$SrTiO_3$ (Nb:STO) heterostructures with varying LSMO-layer thickness, in which the magnetic anisotropy strongly changes depending on the LSMO thickness due to the delicate balance between the strains originating from both the Nb:STO and LAO layers, using x-ray magnetic circular dichroism (XMCD) and photoemission spectroscopy (PES). We successfully detect the clear change of the magnetic behavior of the Mn ions concomitant with the thickness-dependent metal-insulator transition (MIT). Our results suggest that double-exchange interaction induces the ferromagnetism in the metallic LSMO film under tensile strain caused by the $SrTiO_3$ substrate, while superexchange interaction determines the magnetic behavior in the insulating LSMO film under compressive strain originating from the top LAO layer. Based on those findings, the formation of a magnetic dead layer near the LAO/LSMO interface is attributed to competition between the superexchange interaction via Mn $3d_{3z^2-r^2}$ orbitals under compressive strain and the double-exchange interaction via the $3d_{x^2-y^2}$ orbitals.

**Introduction**

Oxide electronics has attracted considerable interest because it has a variety of potential applications both in electronics and spintronics and because it provides us with the opportunity to understand the rich fundamental physics of strongly correlated electron systems [1]. In heterostructures with transition-metal perovskite oxides, the complex interplay among multiple degrees of freedom of the correlated electrons yields unusual quantum phenomena such as superconductivity, colossal magnetoresistance, and high-mobility two-dimensional electron/hole gas [2, 3, 4, 5]. Recently, intriguing spintronics functionality was demonstrated in ferromagnetic-perovskite-oxide (La,Sr)MnO$_3$ (LSMO)-base devices; it was clarified that the magnetic anisotropy of LSMO can be manipulated with extremely low power consumption [6, 7, 8, 9, 10]. For such heterostructures, spin and charge modulations at the interfaces are considered to modify the physical properties of the interfaces from that of bulk materials [11]. The knowledge of the interfacial magnetic and electronic properties of ferromagnetic oxide heterostructures is indispensable for improving the performance of spintronics devices.

The ferromagnetic oxide LSMO is one of the promising materials for spintronics because of its distinctive physical properties such as colossal magnetoresistance, half-metallicity nature, and high Curie temperatures above room temperature [12, 13]. Peculiar physical properties of LSMO including metal-insulator transition (MIT) originate from complex interplay among the charge, spin, and orbital degrees of freedom [14, 15]. In single-crystalline LSMO thin films and heterostructures, the electronic and magnetic properties of LSMO layers are strongly affected by the reduction of dimensionality, epitaxial strain from the substrate or adjacent layers, and/or disorder at the surface or the interface [11, 16, 17]. These external effects lead to the formation of a "magnetic dead layer" near the surface/interface, where the ferromagnetism is suppressed [16, 18]. Previous studies on LSMO thin films under epitaxial strains or with varying thickness have reported that spatial magnetic inhomogeneity leading to phase separation is intimately related to the formation of the magnetic dead layers in LSMO heterostructures [19, 20, 21]. To understand the origin of the magnetic dead layer and the relationship between the epitaxial strain and the formation of the magnetic dead layer, it is important to investigate high-quality crystalline LSMO layers with atomically abrupt interfaces. In particular, the magnetic behavior of ultrathin LSMO layers under epitaxial strain is a key for the application of the ferromagnetic oxide heterostructures for spintronics devices on the nm scale.

In this paper, we investigate the magnetic and electronic states at the interface between LSMO and LaAlO$_3$ (LAO) layers that are epitaxially grown on Nb-doped SrTiO$_3$ (Nb:STO) substrates, namely, LAO/LSMO/NSTO heterostructures with varying thickness of the LSMO layer using x-ray magnetic circular dichroism (XMCD) and photoemission spectroscopy (PES). Here, the magnetic anisotropy of the LSMO layer depends on its thickness due to the delicate balance between strains originating from both the STO and LAO layers. The atomically abrupt interfaces enable us to observe the intrinsic interfacial electronic structure. Our experimental findings demonstrate changes in the magnetic behavior accompanying MIT in the strained ultrathin LSMO layers, providing a mechanism for the formation of the magnetic dead layer at the interfaces in ferromagnetic oxide heterostructures.

**Experimental**
Digitally controlled LAO(2 uc)/LSMO($t$ uc) ($t$ = 5, 10, and 15) heterostructures were epitaxially grown on the atomically flat (001) surfaces of TiO$_2$-terminated Nb-doped SrTiO$_3$ (STO) substrates by molecular-beam epitaxy (MBE). Here, uc denotes a unit cell. The LAO top layer protects the surface from further oxidization. Details of the growth conditions are described elsewhere [9]. The sample structures of the LAO/LSMO/Nb:STO heterostructures are shown in Fig. 1(a). The magnetic properties of the heterostructures were measured using a superconducting quantum interference device (SQUID) magnetometer. The PES and XMCD experiments were performed at the helical undulator beamline BL23SU of SPring-8 [22, 23, 24]. The monochromator resolution $E/\Delta E$ was 10000. For the XMCD measurements, absorption spectra for circularly polarized x rays with the photon helicity parallel ($\mu^+$) and antiparallel ($\mu^-$) to the spin polarization were taken by reversing the photon helicity at each photon energy $h\nu$ and were recorded in the total-electron-yield mode. The $\mu^+$ and $\mu^-$ spectra were taken for both positive and negative applied magnetic fields and were averaged in order to eliminate spurious dichroic signals arising from the slightly different optical paths for the two circular polarizations. External magnetic fields were applied perpendicular to the sample surfaces ([001] direction). For estimation of the integrated values of the XAS spectra at the Mn $L_{2,3}$ edge, hyperbolic tangent functions were subtracted from the spectra as backgrounds. For the PES measurements, the samples were kept at 20 K under an ultra-high vacuum better than $10^{-8}$ Pa. The total energy resolution including temperature broadening was ~150 meV. The position of the Fermi level ($E_F$) was determined by measuring evaporated gold that was electrically in contact with the samples.

**Results and discussion: Magnetic properties**

Figure 2 shows the magnetic-field dependence of the magnetization of the heterostructures measured by the SQUID magnetometer. The out-of-plane linear magnetic susceptibility including the diamagnetic contribution from the substrates for $t = 5$ uc is positive, while that for $t = 15$ uc is negative (see the red curves in Figs. 2(a) and 2(b)). This indicates that there exists considerable amount of paramagnetic Mn ions for $t = 5$ uc. Figures 2(c) and 2(d) show the magnetization curves obtained for $t = 5$ and 15 uc, respectively, after subtracting the abovementioned linear components. It should be noted here that the saturation magnetization for $t = 15$ uc is ~4 $\mu_B$/Mn, nearly the full magnetic moment per Mn atom, suggesting that nearly all the Mn ions contribute to the ferromagnetism. In contrast, the saturation magnetization is much smaller than the full moment for $t = 5$ uc, as shown in Fig. 2(c). For $t = 5$ uc, the out-of-plane susceptibility near the zero field is much larger than the in-plane one, as shown in the inset of Fig. 2(c), indicating the perpendicular magnetic anisotropy. The most important point here is that the easy magnetization axis is changed with $t$, i.e., the film with $t = 5$ uc has perpendicular magnetic anisotropy, while the one with $t = 15$ uc has in-plane magnetic anisotropy.

**Results and discussion: PES**

To reveal the changes of the electronic structure with the LSMO thickness, we have performed x-ray photoemission spectroscopy (XPS) for core levels and resonant photoemission spectroscopy (RPES) for the valence band (VB). Figure 3 shows the Mn $2p$ core-level XPS spectra of the LAO/LSMO heterostructures. Note that a small peak (well-screened peak) appears on the low binding-energy ($E_B$) side of the $2p_{3/2}$ peak only for $t = 15$ uc. This well-screened feature of the Mn $2p$ spectra originates from the screening of the $2p$ core hole by the ferromagnetic metallic electrons in the LSMO layer [25]. The appearance of this peak reflects the increase of the metallicity of the 15-uc LSMO layer. In the previous PES studies of LSMO thin films [25, 26], the well-screened peaks observed in hard x-ray PES with high bulk sensitivity were larger than those in soft x-ray PES probably because of defects/disorder near the surface or interface. Thus, the observation of the well-screened peak with soft x-ray PES in the present study suggests the high-quality crystallinity of the LAO/LSMONb:STO heterostructures with the atomically abrupt interfaces. In addition, if one compares the main peaks of the Mn $2p$ spectra, a core-level shift depending on the thickness of the LSMO layer is identified (see the dashed lines in Fig. 3).

Figures 4(a) and 4(b) show the off-resonance photoemission spectra of the VB taken at $h\nu = 637$ eV for the LAO/LSMO/Nb:STO heterostructures. It should be noted here that there is a gap at the Fermi level ($E_F$) for $t = 5$ uc, while the spectrum for $t = 15$ uc has a finite density of states at $E_F$, that is, metal-insulator transition (MIT) occurs at a thickness between $t = 5$ uc and 15 uc. This is analogous to the thickness-dependent MIT in LSMO layers sandwiched between STO [17] and indicates that the core-level shift observed in the Mn $2p$ spectra reflects the chemical potential shift accompanied with the MIT. To reveal the changes of the Mn $3d$ states across the MIT in details, the Mn $2p$-$3d$ RPES spectra taken at $h\nu = 642.5$ eV are shown in Figs. 4(c) and 4(d). In contrast to the insulting state for $t = 5$ uc, the metallic nature of the Mn $3d$ states for $t = 15$ uc is obvious. This is consistent with the appearance of the well-screened feature observed in the Mn $2p$ spectra for $t = 15$ uc shown in Fig. 3.

**Results and discussion: XMCD**

Figure 5 shows the Mn $L_{2,3}$ XAS and XMCD spectra of the LAO/LSMO/Nb:STO heterostructures measured at a magnetic field $\mu_0 H$ of 1 T applied perpendicular to the films and at a temperature $T = 5$ K. In general, the line shapes of XAS/XMCD spectra reflect the local electronic/magnetic properties. The XAS and XMCD spectra of the 15-uc sample are similar to those of thick LSMO layers on STO substrates [27]. This result indicates that the LSMO layer with $t = 15$ uc in our LAO/LSMO/Nb:STO heterostructure has nearly the same electronic structure as thick LSMO layers under tensile strain reported in previous reports [27, 28, 29, 30]. In contrast, the XAS spectrum of the 5-uc film is different from that of 15 uc film, but is similar to that of thick LSMO layers deposited on LAO substrates [28, 29, 30], suggesting that the 5-uc LSMO layer is compressively strained by the top LAO layer. The strongest XAS peak for $t$ = 15 uc is located at slightly lower energy than that for $t$ = 5 uc, which is similar to the shift of the main peaks observed for LSMO/LAO and LSMO/STO [31]. Note that this peak shift is opposite to the thickness dependence of LSMO/STO [27]. Comparing the line shapes of the XMCD spectra for the different LSMO thicknesses, as shown in Fig. 1(b), the electronic structure of the LSMO layer clearly changes as the thickness decreases from 15 uc to 5 uc.

Figures 6(a) and 6(b) show the $H$ dependence of the XMCD spectra of the LAO/LSMO($t$ = 15 uc and 5 uc)/Nb:STO heterostructures, respectively. The XMCD intensities increase with increasing $H$. As for the XMCD line shapes, they are approximately independent of $H$ for the 15-uc film [inset of Fig. 6(a)], while there are appreciable spectral line shape changes with $H$ around 640 eV for the 5-uc film [inset of Fig. 6(b)]. This means that a paramagnetic and/or antiferromagnetic component exists in the 5-uc film in addition to the ferromagnetic component, while almost all the Mn ions contribute to the ferromagnetism in the 15-uc film. Note that the XMCD intensity for $t$ = 15 uc is exceedingly larger than that for $t$ = 5 uc, consistent with the magnetization measured by SQUID shown in Fig. 2. Figures 6(c) and 6(d) show the $H$- and $T$-dependences of the XMCD intensities of the LAO/LSMO/Nb:STO heterostructures, respectively. As one can see, the magnetic behavior is clearly different between the thin and thick LSMO films. The magnetization curve for $t$ = 5 uc is rapidly saturated in comparison with that for $t$ = 15 uc although the saturation magnetization per Mn ion of the film for $t$ = 5 uc is as small as one third of that for $t$ = 15 uc, as shown in Fig. 6(c). While the magnetization for $t$ = 15 uc rapidly decreases with increasing $T$, that for $t$ = 5 uc only gradually decreases with increasing $T$, as shown in Fig. 6(d). Considering the observations using SQUID shown in Fig. 2, this result likely reflects the change of the magnetic anisotropy depending on $t$. In thick LSMO films grown on STO substrates, double-exchange interaction between the $Mn^{3+}$ and $Mn^{4+}$ ions stabilizes ferromagnetism with an in-plane easy magnetization axis [30, 32]. In our work, these bulk-like magnetic properties are consistently observed in the thick 15-uc film. In contrast, the magnetic behavior of the film for $t = 5$ uc is completely different from that reported for single LSMO films grown on STO [32].

**Discussion: Competition between different magnetic interactions**

The XMCD and RPES results on the LAO/LSMO/Nb:STO heterostructures with varying LSMO-layer thickness indicate the changes of the magnetic behavior accompanied by MIT. For bulk LSMO and thick LSMO films, the ferromagnetism arises from double-exchange interaction [33] between the $Mn^{3+}$ and $Mn^{4+}$ ions, which leads to the colossal magnetoresistance concomitant with the MIT [15]. As shown in Fig. 1(b), under the tensile strain from the STO substrate, the $x^2$-$y^2$ orbital of the $e_g$ states is preferentially occupied by conduction electrons, leading to the magnetic anisotropy with the in-plane [110] (or [100]) easy magnetization axis. In contrast to the nearly full moment in the 15-uc LSMO layer observed by SQUID and XMCD, the saturation magnetization of the 5-uc LSMO film is much smaller than the full moment (see Figs. 2(c), 2(d), 6(c), and 6(d)). This may arise from superexchange interaction that plays an essential role in the magnetism of the insulating phase of the ultrathin LSMO layer, where the $d$ electrons are localized. When the LSMO layer is thin enough in the LAO/LSMO/Nb:STO heterostructure, the compressive strain is predominant because the lattice mismatch of LSMO with LAO of ~3% is considerably larger than that with STO of ~1% [28]. Under the compressive strain from the LAO layer, the $d_{3z^2-r^2}$ orbital of the $Mn^{3+}$ ion is preferentially occupied, as shown in Fig. 1(b), resulting in the magnetic anisotropy with the out-of-plane [001] easy magnetization axis. The superexchange interaction path $Mn^{4+}$-O-$Mn^{3+}$ in the out-of-plane direction involving the $d_{3z^2-r^2}$ orbital is expected to be ferromagnetic, while the same interaction within the plane is antiferromagnetic [34]. It follows from these arguments that the change of the magnetic anisotropy and the reduction of the magnetization accompanying the MIT originate from the compressive strain in the LSMO layer.

Whereas most of the Mn ions contribute to the ferromagnetism in the metallic LSMO layer, only a part of the Mn ions contributes to the ferromagnetism in the insulating LSMO layer. From the *M-H* curve of the heterostructure with $t = 5$ uc shown in Fig. 6(c), the saturation magnetization and the paramagnetic susceptibility $\chi$ in the high magnetic fields ($\mu_0 H > 1$ T) are estimated to be ~0.72 $\mu_B$/Mn and $5.23 \times 10^2$ $\mu_B$/Mn/T, respectively. The saturation magnetization is nearly identical to that estimated from the macroscopic SQUID measurements shown in Fig. 2(c), and indicates that only 25% of the total Mn ions contribute to the ferromagnetism. The rest of Mn ions may be coupled with each other antiferromagnetically through superexchange interaction via $Mn^{4+}$-O-$Mn^{4+}$ or $Mn^{3+}$-O-$Mn^{3+}$ paths. Under the compressive strain, the preferential occupation of the $d_{3z^2-r^2}$ orbital in the LSMO layer favors spin orientation perpendicular to the *ab*-plane [16, 28, 35]. This is evident in the SQUID results shown in Fig. 2(a) and 2(b), where the out-of-plane [001] paramagnetic susceptibility including the diamagnetic contribution from the substrate is positive in the insulating heterostructure for $t$ = 5 uc, but negative in the ferromagnetic heterostructure for $t$ = 15 uc. We note that the in-plane [100] paramagnetic susceptibility is negative in both cases ($t$ = 5 and 15 uc). By analyzing the paramagnetic susceptibility $\chi$ using the Curie-Weiss law $\chi(T) = C/(T - \theta)$, where $C$ is the Curie constant and $\theta$ is the Weiss temperature, under the assumption that 75% of the Mn ions are paramagnetic, the Weiss temperature is estimated approximately to be – 150 K. The negative Weiss temperature is qualitatively consistent with antiferromagnetic superexchange via the $Mn^{3+}$-O-$Mn^{3+}$ path. Therefore, the occupation of the $d_{3z^2-r^2}$ orbital under the compressive strain (see Fig. 1(b)) can explain why the paramagnetic susceptibility appears preferentially for the out-of-plane direction in the insulating heterostructure.

The present experimental findings may provide a key knowledge to understand the formation of a magnetic dead layer near the LAO/LSMO interface. In the metallic LSMO case, when the interface is not ideal and has extrinsic defects such as atomic intermixing and structural deformation [11, 29], the conduction electrons are scattered by these defects/disorder, resulting in the weakening of the double-exchange interaction [36]. In addition to the compressive strain, oxygen octahedral rotation (OOR) for the LSMO layers is induced by the top LAO layer [9, 37, 38]. The OOR slightly reduces the in-plane bond angle of the Mn-O-Mn path in the vicinity of the interface, and weakens the double-exchange interaction. As a consequence, competition between the double-exchange and superexchange interaction depending on strain leads to the formation of the magnetic dead layer near the interface.

For the origin of the magnetic dead layer at the interfaces of LSMO thin films under tensile strain, phase-separation models have been proposed [19, 20, 21]. In a previous XMCD study on thickness-dependent MIT in a LSMO/STO heterostructure [27], it has been revealed that paramagnetic, superparamagnetic, and ferromagnetic phases co-exist in the magnetic dead layer. The magnetic inhomogeneity or the phase separation is probably associated with the spatial fluctuation of hole distribution and superexchange interaction in the LSMO layer. In the present case of the insulating LSMO film under compressive strain, it is likely that the ferromagnetic and super-paramagnetic phases arise from the ferromagnetic $Mn^{4+}$-O-$Mn^{3+}$ superexchange path and that the rest of Mn ions contribute to the paramagnetism with antiferromagnetic coupling. Since the antiferromagnetic superexchange interaction of the $Mn^{3+}$-O-$Mn^{3+}$ path is predominant in the area of sparse hole carriers, the Weiss temperature becomes negative. In the insulating LSMO heterostructure, the thickness of the magnetic dead layer is determined by the strain-relaxed region near the heterointerface because the magnetic anisotropy due to the superexchange interaction strongly depends on strain.

Finally, the experimental findings that the changes of the magnetic behavior concomitant with MIT may suggest applications of ferromagnetic oxide interfaces. If one can control the position of $E_F$ in the LSMO layer by applying bias voltage, the magnetic behavior will change depending on the hole concentration in the LSMO layer. This may be analogous to the orbital-controlled magnetization switching observed in an LSMO/STO/LSMO magnetic tunnel junction [10]. Additionally, based on the present XMCD and SQUID results, the magnetic behavior observed for $t = 5$ uc reflects the LSMO layer under compressive strain. It should be noted here that the top LAO layer significantly affects the magnetic behavior in the insulating LSMO layer even though the thickness of the LAO layer is merely 2-uc thick. This indicates that the structural design of the ferromagnetic oxide heterostructures on the nm scale possibly controls the performance or properties of such devices.

**Conclusion**

In epitaxially grown single-crystalline $LaAlO_3$(2 uc)/$La_{0.6}Sr_{0.4}MnO_3$($t$ uc)/Nb:$SrTiO_3$(001) heterostructures with varying LSMO-layer thickness $t$, the magnetic anisotropy strongly changes depending on the LSMO thickness due to the delicate balance between strains originating from both the Nb:STO and LAO layers. The film with $t = 5$ uc has perpendicular magnetic anisotropy, while the one with $t = 15$ us has in-plane magnetic anisotropy. To understand the interfacial magnetic and electronic properties of the LAO/LSMO/Nb:STO heterostructures with different LSMO thicknesses, we have performed XMCD and PES measurements. The observation of the well-screened features in the Mn $2p$ XPS spectra suggests a significantly improved crystal structure quality at the interface between the LAO and LSMO layers in the MBE-grown thin films. The Mn $L_{2,3}$ XMCD and Mn $2p$-$3d$ RPES demonstrate the change of the magnetic behavior accompanied by MIT. The magnetic behavior of the insulating thin LSMO layer originates from superexchange interaction between the Mn ions under the compressive strain from the top LAO layer, while double-exchange interaction is predominant in the ferromagnetic metallic LSMO layer. Based on the present experimental findings, the formation of the magnetic dead layer is attributed to competition between the superexchange and double-exchange interaction in the strain-relaxed region near the heterointerface. It is likely that the changes of the magnetic behavior concomitant with MIT provide key aspects of ferromagnetic oxide heterostructures for the structural design in nm scale and the device application.

**Acknowledgments**

This work was supported by Grants-in-Aid for Scientific Research (18H03860 and 19K21960) from the Japan Society for the Promotion of Science (JSPS) and PRESTO Program (JPMJPR19LB, JPMJCR1777) of Japan Science and Technology Agency (JST), Japan. This work was partially supported the Spintronics Research Network of Japan (Spin-RNJ). This work was performed under the Shared Use Program of Japan Atomic Energy Agency (JAEA) Facilities (Proposal No. 2017B-E19 and 2020A-E18) supported by JAEA Advanced Characterization Nanotechnology Platform as a program of "Nanotechnology Platform" of the Ministry of Education, Culture, Sports, Science and Technology (MEXT) (Proposal No. A-17-AE-0038 and A-20-AE-0018). The experiment at SPring-8 was approved by the Japan Synchrotron Radiation Research Institute (JASRI) Proposal Review Committee (Proposal No. 2017B3841 and 2020A3841).

**References**

[1] H. Y. Hwang, Y. Iwasa, M. Kawasaki, B. Keimer, N. Nagaosa, and Y. Tokura, Emergent phenomena at oxide interfaces, Nat. Mater. **11**, 103 (2012).

[2] A. Ohtomo, D. A. Muller, J. L. Grazul, and H. Y. Hwang, Artificial charge-modulationin atomic-scale perovskite titanate superlattices, Nature **419**, 378 (2002).

[3] H. Takagi and H. Y. Hwang, An Emergent Change of Phase for Electronics, Science **327**, 1601 (2010).

[4] J. Mannhart and D. G. Schlom, Oxide Interfaces–An Opportunity for Electronics, Science **327**, 1607 (2010).

[5] L. D. Anh, S. Kaneta, M. Tokunaga, M. Seki, H. Tabata, M. Tanaka, and S. Ohya, High-Mobility 2D Hole Gas at a $SrTiO_3$ Interface, Adv. Mater. **32**, 1906003 (2020).

[6] M. Bowen, M. Bibes, A. Barthélémy, J.-P. Contour, A. Anane, Y. Lemaître, and A. Fert, Nearly total spin polarization in $La_{2/3}Sr_{1/3}MnO_3$ from tunneling experiments, Appl. Phys. Lett. **82**, 233 (2003).

[7] S. M.Wu, S. A. Cybart, P. Yu, M. D. Rossell, J. X. Zhang, R. Ramesh, and R. C. Dynes, Reversible electric control of exchange bias in a multiferroic field-effect device, Nat. Mater. **9**, 756 (2010).

[8] R. Werner, A. Yu. Petrov, L. Alvarez Miño, R. Kleiner, D. Koelle, and B. A. Davidson, Improved tunneling magnetoresistance at low temperature in manganite junctions grown by molecular beam epitaxy, Appl. Phys. Lett. **98**, 162505 (2011).

[9] L. D. Anh, N. Okamoto, M. Seki, H. Tabata, M. Tanaka, and S. Ohya, Hidden peculiar magnetic anisotropy at the interface in a ferromagnetic perovskite-oxide heterostructure, Sci. Rep. **7**, 8715 (2017).

[10] L. D. Anh, T. Yamashita, H. Yamasaki, D. Araki, M. Seki, H. Tabata, M. Tanaka, and S. Ohya, Ultralow-Power Orbital-Controlled Magnetization Switching Using a Ferromagnetic Oxide Interface, Phys. Rev. Applied **12**, 041001 (2019).

[11] L. Fitting Kourkoutis, J. H. Song, H. Y. Hwang, and D. A. Muller, Microscopic origins for stabilizing room-temperature ferromagnetism in ultrathin manganite layers, PNAS **107**, 11682 (2010).

[12] A. Urushibara, Y. Moritomo, T. Arima, A. Asamitsu, G. Kido, and Y. Tokura, Insulator-metal transition and giant magnetoresistance in $La_{1-x}Sr_xMnO_3$, Phys. Rev. B **51**, 14103 (1995).

[13] J.-H. Park, E. Vescovo, H.-J. Kim, C. Kwon, R. Ramesh, and T. Venkatesan, Direct

evidence for a half-metallic ferromagnet, Nature **392**, 794 (1998).

[14] M. Imada, A. Fujimori, and Y. Tokura, Metal-insulator transitions, Rev. Mod. Phys. **70**, 1039 (1998).

[15] Y. Tokura and Y. Tomioka, Colossal magnetoresistive manganites, J. Magn. Magn. Mater. **200**, 1 (1999).

[16] Y. Konishi, Z. Fang, M. Izumi, T. Manako, M. Kasai, H. Kuwahara, M. Kawasaki, K. Terakura, and Y. Tokura, Orbital-State-Mediated Phase-Control of Manganites, J. Phys. Soc. Jpn. **68**, 3790 (1999).

[17] K. Yoshimatsu, K. Horiba, H. Kumigashira, E. Ikenaga, and M. Oshima, Thickness dependent electronic structure of $La_{0.6}Sr_{0.4}MnO_3$ layer in $SrTiO_3/La_{0.6}Sr_{0.4}MnO_3/SrTiO_3$ heterostructures studied by hard x-ray photoemission spectroscopy, Appl. Phys. Lett. **94**, 071901 (2009).

[18] M. Izumi, Y. Ogimoto, T. Manako, M. Kawasaki, and Y. Tokura, Interface Effect and Its Doping Dependence in $La_{1-x}Sr_xMnO_3/SrTiO_3$ Superlattices, J. Phys. Soc, Jpn. **71**, 2621 (2002).

[19] Y. H. Sun, Y. G. Zhao, H. F. Tian, C. M. Xiong, B. T. Xie, M. H. Zhu, S. Park, W. Wu, J. Q. Li, and Q. Li, Electric and magnetic modulation of fully strained dead layers in $La_{0.67}Sr_{0.33}MnO_3$ films, Phys. Rev. B **78**, 024412 (2008).

[20] Y. Zhu, K. Du, J. Niu, L. Lin, W. Wei, H. Liu, H. Lin, K. Zhang, T. Yang, Y. Kou, J. Shao, X. Gao, X. Xu, X. Wu, S. Dong, L. Yin, and J. Shen, Chemical ordering suppresses large-scale electronic phase separation in doped manganites, Nat. Comm. **7**, 11260 (2016).

[21] G. Shibata, K. Yoshimatsu, E. Sakai, K. Ishigami, S. Sakamoto, Y. Nonaka, F.-H. Chang, H.-J. Lin, D.-J. Huang, C.-T. Chen, H. Kumigashira, and A. Fujimori, Temperature Evolution of Magnetic Phases Near the Thickness-Dependent Metal–Insulator Transition in $La_{1-x}Sr_xMnO_3$ Thin Films Observed by XMCD, JPS Conf. Proc. **30**, 011072 (2020).

[22] A. Yokoya, T. Sekiguchi, Y. Saitoh, T. Okane, T. Nakatani, T. Shimada, H. Kobayashi, M. Takao, Y. Teraoka, Y. Hayashi, S. Sasaki, Y. Miyahara, T. Harami, and T. A. Sasaki, Soft X-ray Beamline Specialized for Actinides and Radioactive Materials Equipped with a Variably Polarizing Undulator, J. Synchrotron Rad. **5**, 10 (1998).

[23] J. Okamoto, K. Mamiya, S.-I. Fujimori, T. Okane, Y. Saitoh, Y. Muramatsu, A. Fujimori, S. Ishiwata, and M. Takano, Magnetic Circular X-ray Dichroism Study of Paramagnetic and Anti-Ferromagnetic States in $SrFeO_3$ Using a 10-T Superconducting

Magnet, AIP Conf. Proc. **705**, 1110 (2004).

[24] Y. Saitoh, Y. Fukuda, Y. Takeda, H. Yamagami, S. Takahashi, Y. Asano, T. Hara, K. Shirasawa, M. Takeuchi, T. Tanaka, and H. Kitamura, Performance upgrade in the JAEA actinide science beamline BL23SU at SPring-8 with a new twin-helical undulator, J. Synchrotron Rad. **19**, 388 (2012).

[25] K. Horiba, M. Taguchi, A. Chainani, Y. Takata, E. Ikenaga, D. Miwa, Y. Nishino, K. Tamasaku, M. Awaji, A. Takeuchi, M. Yabashi, H. Namatame, M. Taniguchi, H. Kumigashira, M. Oshima, Lippmaa, M. Kawasaki, H. Koinuma, K. Kobayashi, T. Ishikawa, and S. Shin, Nature of the Well Screened State in Hard X-Ray Mn $2p$ Core-Level Photoemission Measurements of $La_{1-x}S_xMnO_3$ Films, Phys. Rev. Lett. **93**, 236401 (2004).

[26] T. Pincelli, V. Lollobrigida, F. Borgatti, A. Regoutz, B. Gobaut, C. Schlueter, T.-L. Lee, D. J. Payne, M. Oura, K. Tamasaku, A. Y. Petrov, P. Graziosi, F. Miletto Granozio, M. Cavallini, G. Vinai, R. Ciprian, C. H. Back, G. Rossi, M. Taguchi, H. Daimon, G. van der Laan, and G. Panaccione, Quantifying the critical thickness of electron hybridization in spintronics materials, Nat. Comm. **8**, 16051 (2017).

[27] G. Shibata, K. Yoshimatsu, E. Sakai, V. R. Singh, V. K. Verma, K. Ishigami, T. Harano, Tl. Kadono, Y. Takeda, T. Okane, Y. Saitoh, H. Yamagami, A. Sawa, H. Kumigashira, M. Oshima, T. Koide, and A. Fujimori, Thickness-dependent ferromagnetic metal to paramagnetic insulator transition in $La_{0.6}Sr_{0.4}MnO_3$ thin films studied by x-ray magnetic circular dichroism, Phys. Rev. B **89**, 235123 (2014).

[28] C. Aruta, G. Ghiringhelli, V. Bisogni, L. Braicovich, N. B. Brookes, A. Tebano, and G. Balestrino, Orbital occupation, atomic moments, and magnetic ordering at interfaces of manganite thin films, Phys. Rev. B **80**, 014431 (2009).

[29] S. Valencia, L. Paña, Z. Konstantinovic, Ll. Balcells, R. Galceran, D. Schmitz, F. Sandiumenge, M. Casanove, and B. Martínez, Intrinsic antiferromagnetic/insulating phase at manganite surfaces and interfaces, J. Phys.: Condens. Matter **26**, 166001 (2014).

[30] G. Shibata, M. Kitamura, M. Minohara, K. Yoshimatsu, T. Kadono, K. Ishigami, T. Harano, Y. Takahashi, S. Sakamoto, Y. Nonaka, K. Ikeda, Z. Chi, M. Furuse, S. Fuchino, M. Okano, J.-i. Fujihira, A. Uchida, K. Watanabe, H. Fujihira, S. Fujihira, A. Tanaka, H. Kumigashira, T. Koide, and A. Fujimori, Colossal electromagnon excitation in the non-cycloidal phase of $TbMnO_3$ under pressure, npj Quantum Mater. **3**, 3 (2018).

[31] B. Cui, S. Song, F. Li, G. Y. Wang, H. J. Mao, J. J. Peng, F. Zeng, and F. Pan, Tuning

the entanglement between orbital reconstruction and charge transfer at a film surface, Sci. Rep. **4**, 4206 (2014).

[32] F. Tsui, M. C. Smoak, T. K. Nath, and C. B. Eom, Strain-dependent magnetic phase diagram of epitaxial $La_{0.67}Sr_{0.33}MnO_3$ thin films, Appl. Phys. Lett. **76**, 2421 (2000).

[33] C. Zener, Interaction between the *d*-Shells in the Transition Metals. II. Ferromagnetic Compounds of Manganese with Perovskite Structure, Phys. Rev. **82**, 403 (1951).

[34] J. B. Goodenough, A. Wold, R. J. Arnott, and N. Menyuk, Relationship Between Crystal Symmetry and Magnetic Properties of Ionic Compounds Containing $Mn^{3+}$, Phys. Rev. **124**, 373 (1961).

[35] T. K. Nath, R. A. Rao, D. Lavric, and C. B. Eom, Effect of three-dimensional strain states on magnetic anisotropy of $La_{0.8}Ca_{0.2}MnO_3$ epitaxial thin films, Appl. Phys. Lett. **74**, 1615 (1999).

[36] Y. Feng, K-J. Jin, L. Gu, X. He, C. Ge, Q.-H. Zhang, M. He, Q.-L. Guo, Q. Wan, M. He, H.-B. Lu, and G. Yang, Insulating phase at low temperature in ultrathin $La_{0.8}Sr_{0.2}MnO_3$ films, Sci. Rep. **6**, 22382 (2016).

[37] C. L. Jia, S. B. Mi, M. Faley, U. Poppe, J. Schubert, and K. Urban, Oxygen octahedron reconstruction in the $SrTiO_3$/$LaAlO_3$ heterointerfaces investigated using aberration-corrected ultrahigh-resolution transmission electron microscopy, Phys. Rev. B **79**, 081405(R) (2009).

[38] T. T. Fister, H. Zhou, Z. Luo, S. S. A. Seo, S. O. Hruszkewycz, D. L. Proffit, J. A. Eastman, P. H. Fuoss, P. M. Baldo, H. N. Lee, and D. D. Fong, Octahedral rotations in strained $LaAlO_3$/$SrTiO_3$ (001) heterostructures, APL Mater. **2**, 021102 (2014).

**Figures and captions**

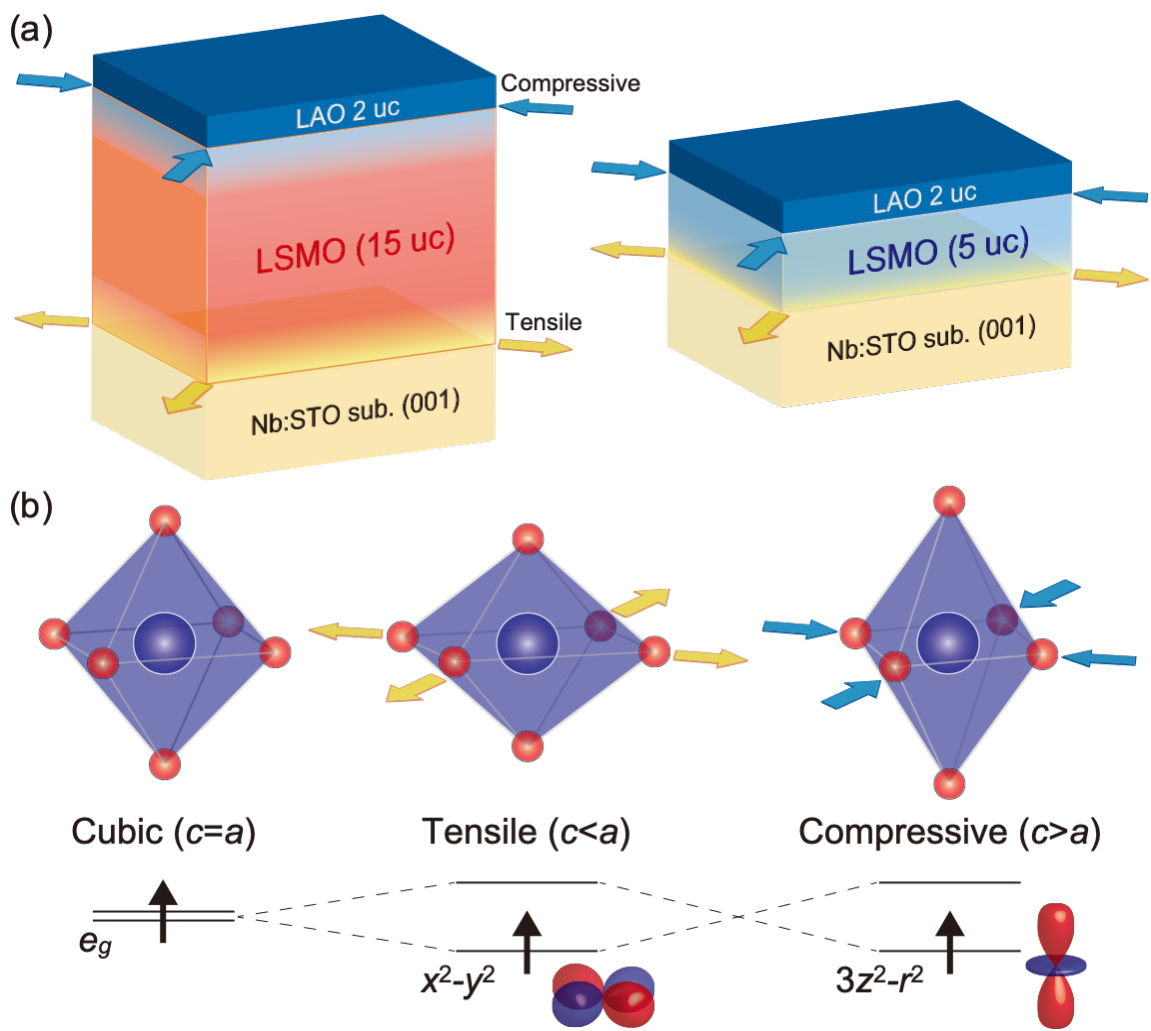

Fig. 1. Schematic pictures of LAO/LSMO/Nb:STO heterostructures. (a) Sample structures of LAO/LSMO/Nb:STO heterostructures with thicknesses of 15 uc (left) and 5 uc (right). Arrows denote the directions of the strains. (b) $Mn^{3+}O_6$ cluster with and without strains. The energy diagram of the 3d $e_g$ states under tensile and compressive strains also shown.

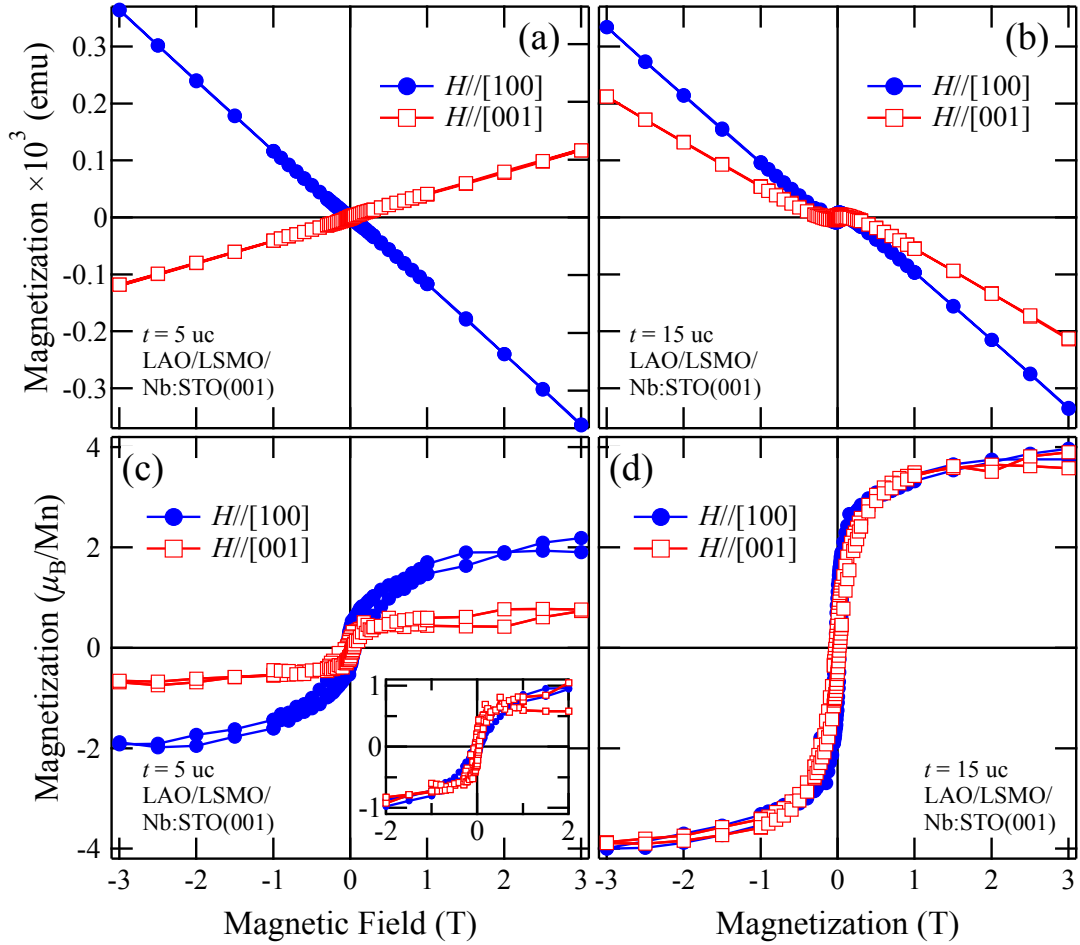

Fig. 2. *M-H* curves of the LAO/LSMO/Nb:STO heterostructures measured by SQUID. (a), (b) Raw *M-H* curves of the 5-uc and 15-uc samples, respectively. The magnetic fields are applied to parallel to the in-plane [100] and out-of-plane [001] directions. (c), (d) *H* dependence of the ferromagnetic components of the 5-uc and 15-uc samples, respectively. Here, the linear components have been subtracted from the *M-H* curves. The inset of (c) shows the normalized magnetization curves.

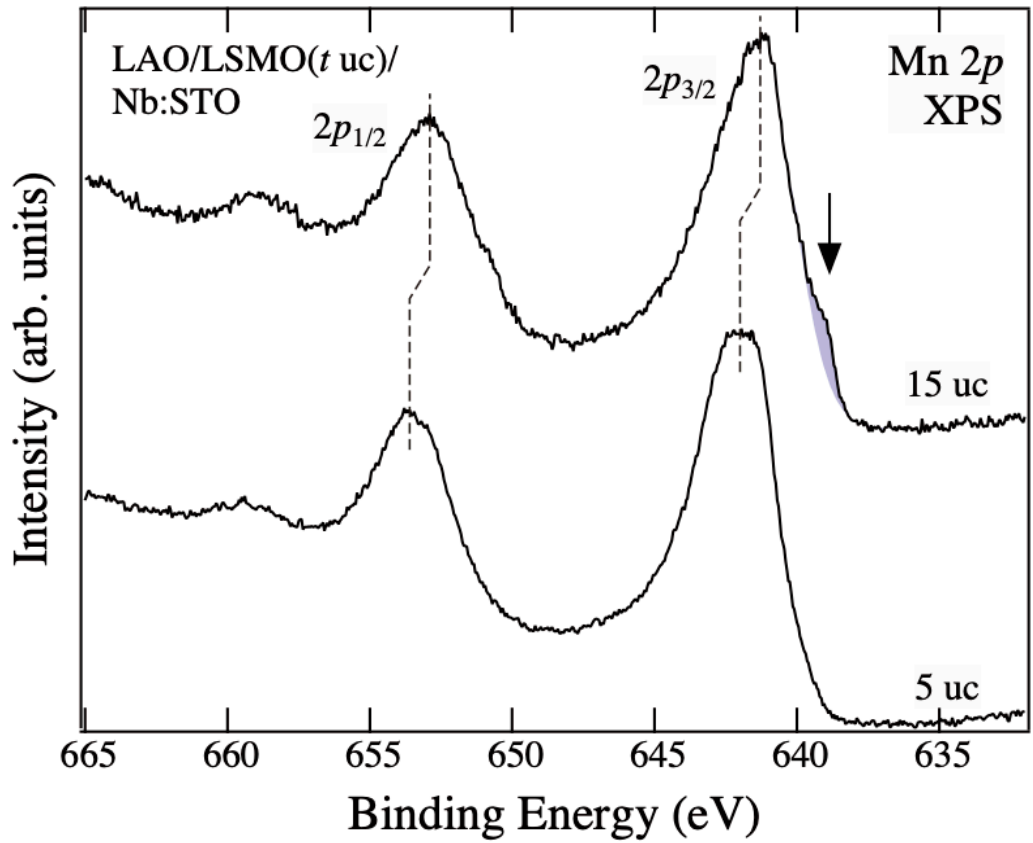

Fig. 3. Mn 2p XPS spectra of the LAO/LSMO/Nb:STO heterostructures. Here, the incident photon energy is $h\nu$ = 1200 eV. The spectra are normalized to the maximum height and a vertical offset put on the spectrum of the film for 15 uc. Shaded area denoted by an allow is the well-screened peak. Vertical dashed lines are guides to the eyes.

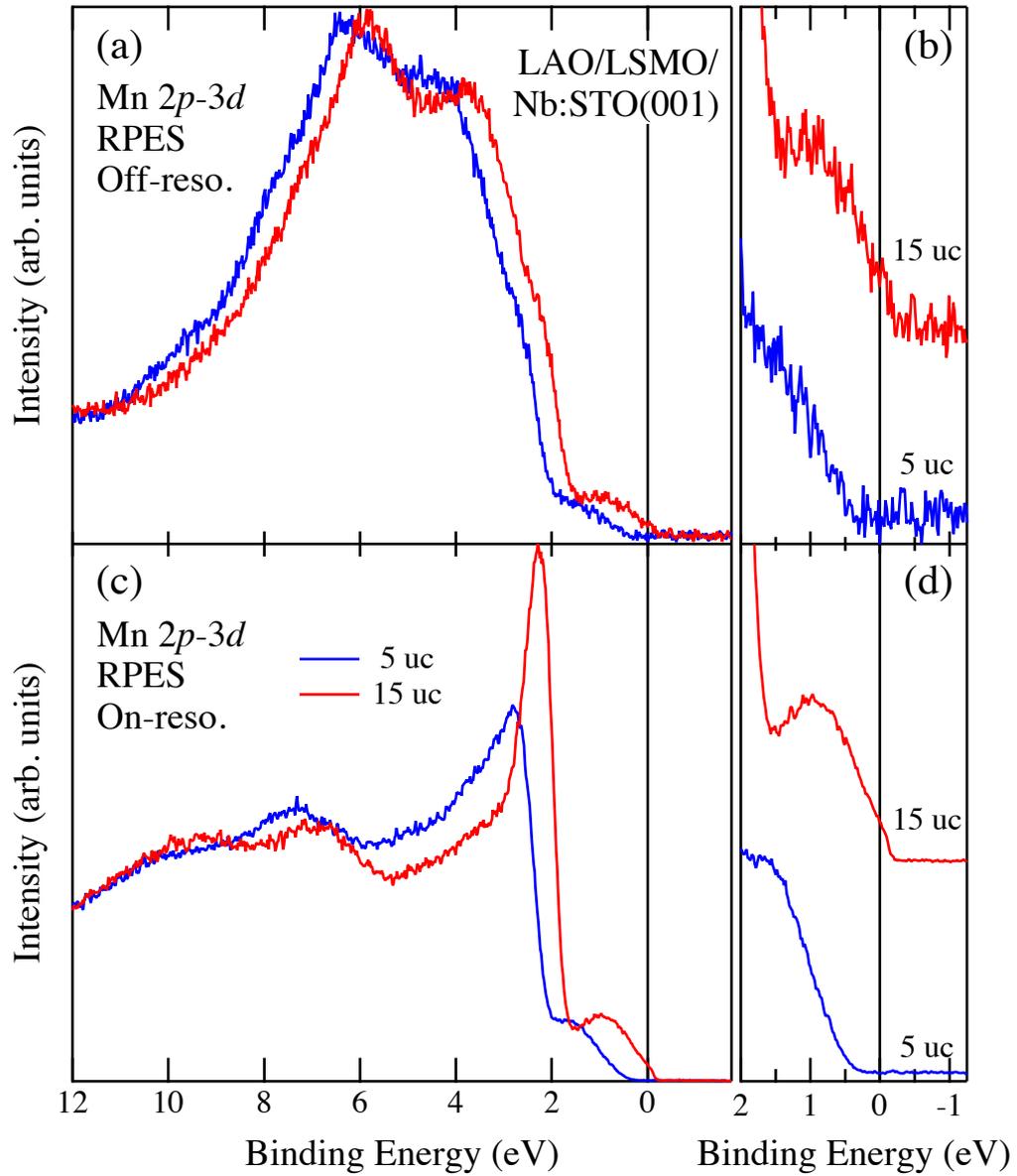

Fig. 4. Mn 2p-3d RPES of the LAO/LSMO/Nb:STO heterostructures. (a) Off-resonance spectra taken at $h\nu$ = 637 eV (at the black arrow shown in Fig. 1). The spectra are normalized to the maximum height. (b) Enlarged plot near $E_F$ of the off-resonance spectra. (c) On-resonance spectra taken at $h\nu$ = 642.5 eV (at the red arrow shown in Fig. 1). (d) Enlarged plot near $E_F$ of the on-resonance spectra. The on-resonance spectra are normalized to the background heights around $E_B$~12 eV.

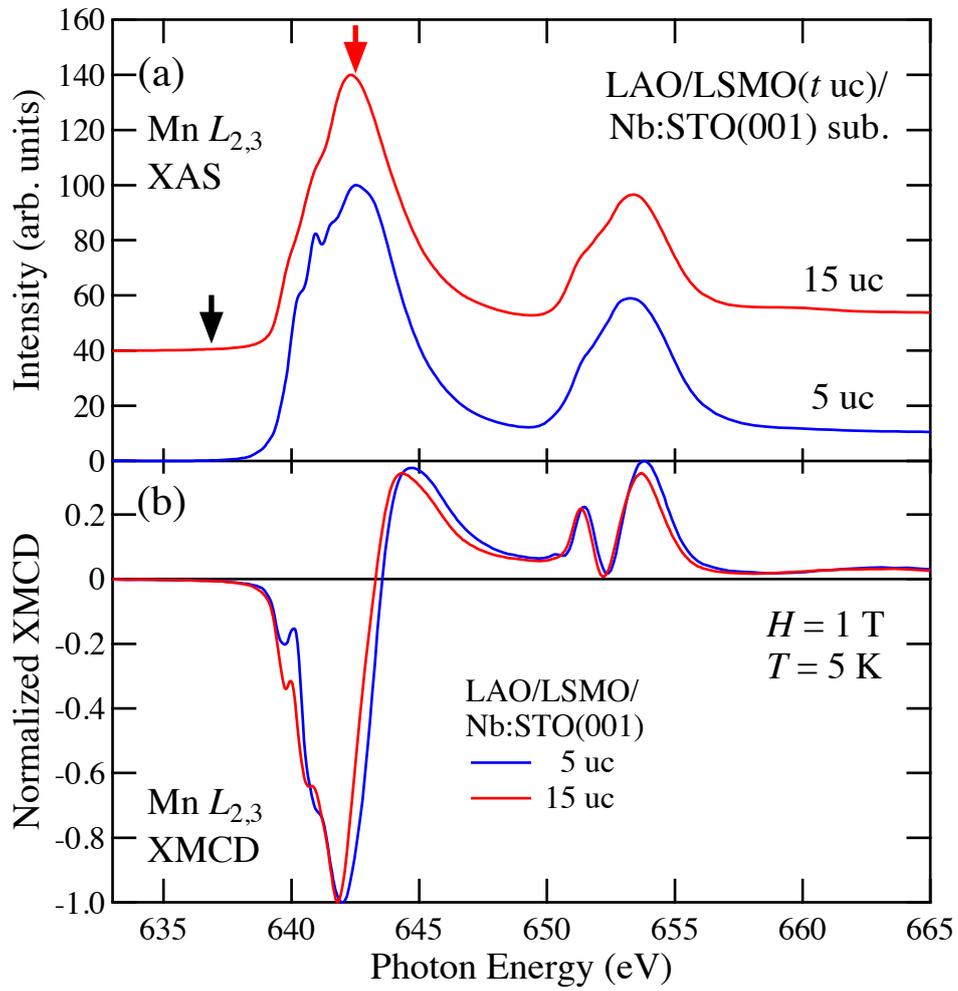

Fig. 5. Mn $L_{2,3}$-edge absorption spectra of the LAO/LSMO/Nb:STO heterostructures. (a) XAS spectra. The intensities are normalized to the maximum height as 100. Red and black arrows denote excitation energies for the off-resonance (637 eV) and on-resonance (642.5 eV) spectra, respectively. (b) XMCD spectra. The spectra have been measured at $\mu_0 H = 1$ T and $T = 5$ K. To compare the spectral line-shapes, the spectra are normalized to the minimum intensity.

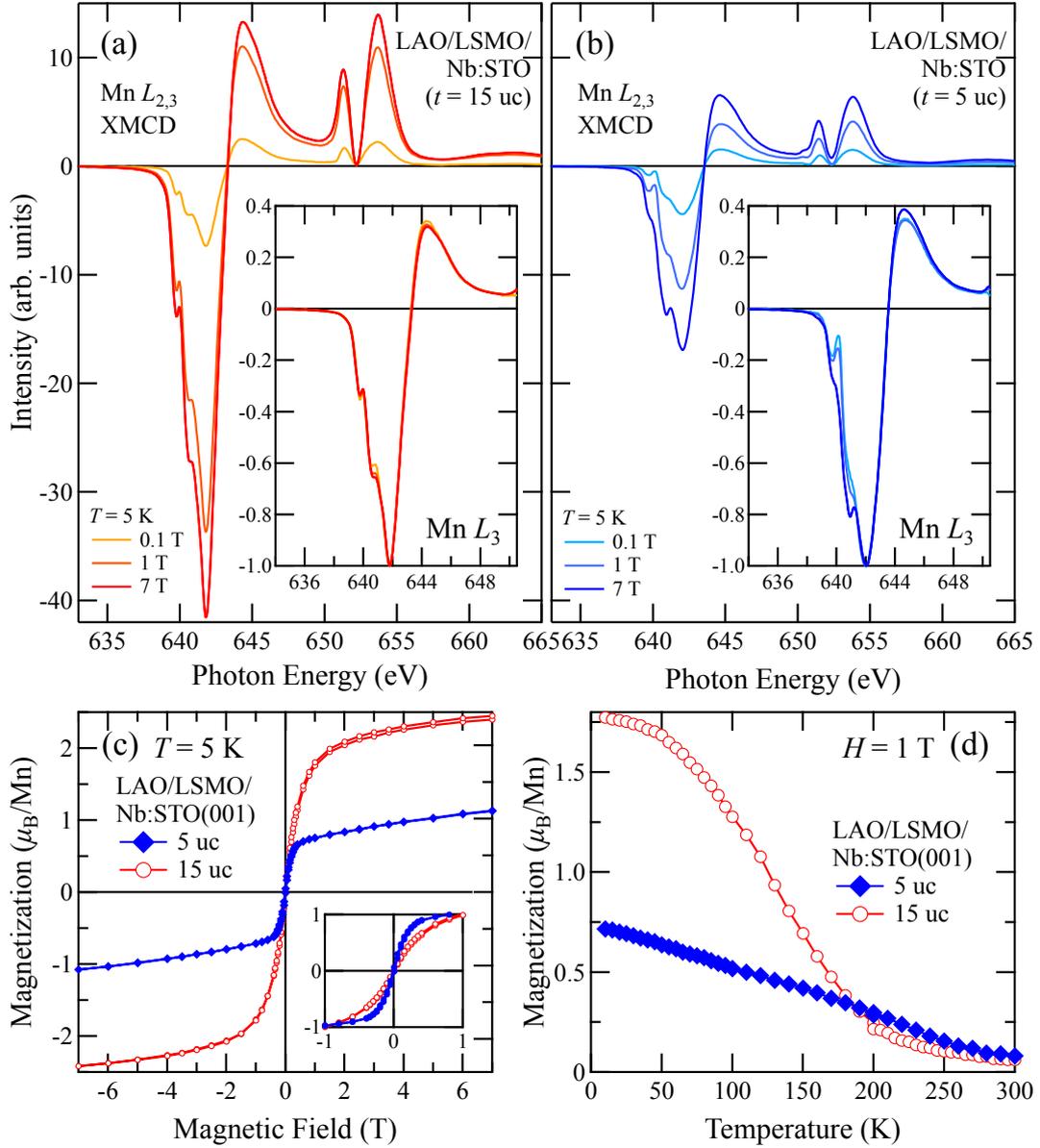

Fig. 6. Magnetic-field and temperature dependences of the Mn $L_{2,3}$ XMCD spectra of the LAO/LSMO/Nb:STO heterostructures. Here, the magnetic field is applied perpendicular to the films. (a), (b) $H$ dependence of XMCD spectra measured at $T$ = 5 K for the LSMO films with thicknesses of 15 uc, and 5 uc, respectively. The insets show the XMCD spectra normalized to the minimum intensity. (c) $H$ dependence of the magnetizations at $T$ = 5 K estimated using the XMCD sum rules. The inset shows the same data normalized at $\mu_0 H$ = 1 T. (d) $T$ dependence of the magnetization at $\mu_0 H$ = 1 T.